\documentclass[runningheads]{llncs}
\usepackage{graphicx}
\usepackage{subcaption}
\usepackage{tabularx}
\usepackage{tikz}
\usetikzlibrary{positioning}

\usepackage{gensymb}
\begin{document}
\title{Shape-based pose estimation for automatic standard views of the knee}
%
%
\author{Lisa Kausch\inst{1,2,3} \and
Sarina Thomas\inst{4} \and
Holger Kunze\inst{5} \and
Jan Siad El Barbari\inst{6} \and
Klaus Maier-Hein\inst{1,2,3}
}
%
\authorrunning{L. Kausch}
%
\institute{Division of Medical Image Computing, German Cancer Research Center (DKFZ), Heidelberg, Germany\\
\email{l.kausch@dkfz-heidelberg.de}\\
\and
National Center for Tumor Diseases (NCT) Heidelberg, Germany\\
\and
Pattern Analysis and Learning Group, Department of Radiation Oncology, Heidelberg University Hospital, Germany\\
\and
Department of Informatics, University of Oslo, Norway\\
\and
Advanced Therapy Systems Division, Siemens Healthineers, Erlangen, Germany\\
\and
MINTOS Research Group, Trauma Surgery Clinic Ludwigshafen, Germany
}
\maketitle              
\begin{abstract}
Surgical treatment of complicated knee fractures is guided by real-time imaging using a mobile C-arm. Immediate and continuous control is achieved via 2D anatomy-specific standard views that correspond to a specific C-arm pose relative to the patient positioning, which is currently determined manually, following a trial-and-error approach at the cost of time and radiation dose. The characteristics of the standard views of the knee suggests that the shape information of individual bones could guide an automatic positioning procedure, reducing time and the amount of unnecessary radiation during C-arm positioning. To fully automate the C-arm positioning task during knee surgeries, we propose a complete framework that enables (1) automatic laterality and standard view classification and (2) automatic shape-based pose regression toward the desired standard view based on a single initial X-ray. A suitable shape representation is proposed to incorporate semantic information into the pose regression pipeline. The pipeline is designed to handle two distinct standard views simultaneously. Experiments were conducted to assess the performance of the proposed system on 3528 synthetic and 1386 real X-rays for the a.-p. and lateral standard. The view/laterality classificator resulted in an accuracy of 100\%/98\% on the simulated and 99\%/98\% on the real X-rays. The pose regression performance was $d\theta_{a.-p}=5.8\pm3.3\degree,\,d\theta_{lateral}=3.7\pm2.0\degree$ on the simulated data and $d\theta_{a.-p}=7.4\pm5.0\degree,\,d\theta_{lateral}=8.4\pm5.4\degree$ on the real data outperforming intensity-based pose regression. 

\end{abstract}
\section{Introduction}
Intraoperative imaging employing a mobile C-arm enables immediate and continuous control during orthopedic and trauma interventions. For optimal fracture reduction and implant placement, correct acquisition of standard views that correspond to a specific C-arm pose relative to the patient is essential \cite{norris1999intraoperative}. Incorrect standard views can exhibit superimposed anatomical structures, leading to overlooked errors that can result in malunion of fractures, functional impairment, or require revision surgeries. To enable deducing all three dimensions of the trauma case, at least two 2D fluoroscopic views are acquired in two distinct planes usually at right angles to each other. The current manual C-arm positioning procedure results in only $20\%$ surgically relevant acquisitions while the remaining $80\%$ are caused by the iterative positioning process, exposing patients and clinical staff to unnecessary radiation \cite{matthews2007navigating}.\\
 \begin{figure}[t]
     \centering
     \includegraphics[width=\textwidth]{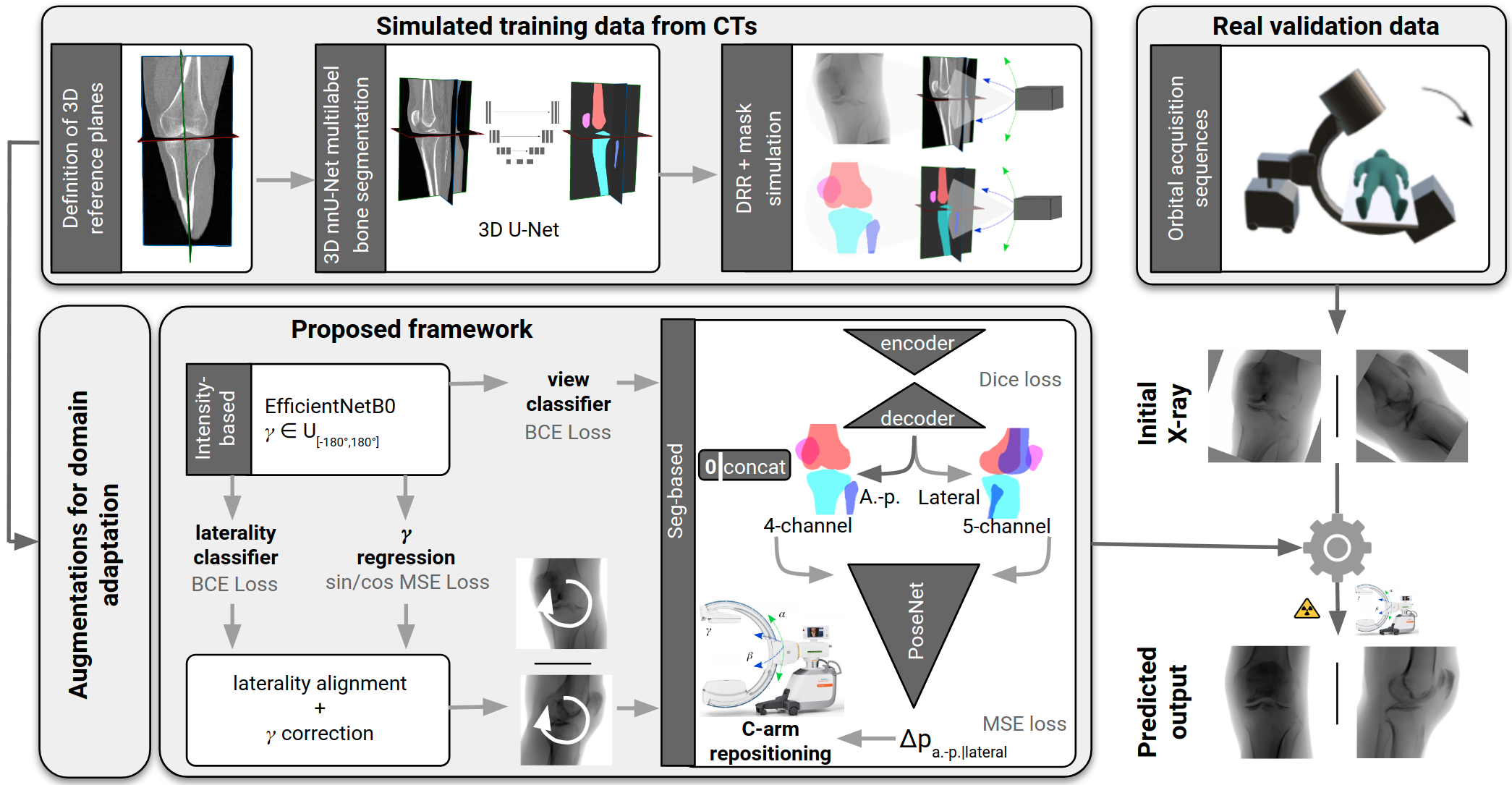}
     \caption{Proposed shape-based pose estimation framework for automatic acquisition of standard views. A single architecture is trained for the representation of 2 distinct standard views simultaneously. The 2-step pipeline consists of a direct intensity-based combined view classification and in-plane rotation regression, followed by a segmentation-based pose regression focusing on the out-of-plane rotation. The pipeline is solely trained on synthetic data with automatically generated ground truth annotations and evaluated on real X-rays.}
     \label{fig:overview}
 \end{figure}
Recent developments towards robotic C-arms ask for automatic positioning methods. Many state of the art approaches require a patient-specific CT for intraoperative real-time simulation \cite{de2017c,fallavollita2014desired,haiderbhai2019automatic} or 2D-3D registration \cite{bott2011use,gong2014cost,miao2018dilated}, external tracking equipment \cite{fotouhi2019interactive,matthews2007navigating}, manual landmark annotation \cite{binder2006image} or do not estimate an optimal pose but reproduce intraoperatively recorded C-arm views employing augmented reality \cite{fotouhi2019interactive,unberath2018augmented}. The inherent prior assumptions and severe inference with the clinical workflow hinder broad clinical applicability until today. In contrast to the majority of anatomical regions, the standard planes of the knee are not orthogonal to each other. The a.-p. standard view is characterized by symmetric projection of the joint gap, femoral and tibial condyles. The tibia surface projects line-shaped and the medial half of the fibula head is superimposed by the tibia. In the lateral standard view, both femoral condyles are aligned and the joint gap is maximized. Automatic deep learning-based positioning for standard views involves specific challenges for image understanding due to overlapping anatomical structures, the presence of surgical implants, and changing viewing directions and showed to benefit from extracting semantic information \cite{kausch2021Carm}. Inspired by that and considering that standard views of the knee anatomy are characterized by the shape information of the individual projected bones, we propose a complete framework to fully automate the C-arm positioning tasks during knee surgeries (Fig.\,\ref{fig:overview}). Our contribution is 4-fold: (1) We propose a novel framework that enables simultaneous automatic standard view classification, laterality classification, in-plane rotation correction, and subsequent view-independent shape-based C-arm positioning to the desired standard view while requiring only a single initial X-ray projection. One pose regression network can handle two distinct standard views of the knee anatomy. A suitable segmentation representation for the knee anatomy is proposed to recognize correct standard views, which explicitly incorporates semantic information to reflect on the actual clinical decision-making of surgeons. Since intraoperative X-rays with reference pose annotations do not exist, the proposed framework is solely trained on simulated data with automatically generated pose annotations. 
(2) We show that the proposed approach outperforms view-specific shape-based and intensity-based pose regression. (3) We show that the proposed shape representation and augmentation strategies aid generalization from simulated training data to real cadaver X-rays. (4) We investigate the importance of individual knee bones on the overall positioning performance for two distinct standard views.

\section{Materials and Methods}
An overview of the complete framework for fully automated C-arm positioning towards desired standard views during knee surgeries is given in Fig.\,\ref{fig:overview}. The anterior-posterior (a.-p.) and the lateral standard view showed to be sufficient for various diagnostic entities \cite{cockshott1985use}.

\subsection{Training data simulation}
To address the interventional data scarcity problem, simulated training data was generated from a collection of CT and C-arm volumes using a realistic DRR simulation framework \cite{unberath2018deepdrr} complemented with corresponding 2D segmentations. Preprocessing involved the following steps:
\begin{figure}[t]
    \centering
    \begin{subfigure}{0.6\textwidth}
        \centering
        \includegraphics[width=0.5\textwidth]{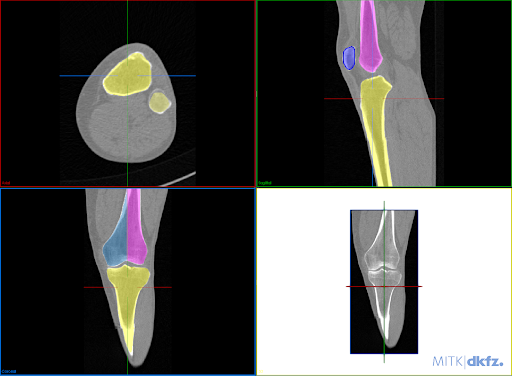}
        \caption{Separation of femur condyles in segmentation}
        \label{fig:seg_repr}
    \end{subfigure}
    \hfill
    \begin{subfigure}{0.31\textwidth}
        \centering
        \includegraphics[width=0.52\textwidth]{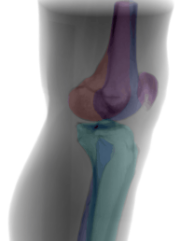}
        \caption{Exemplary simulation}
        \label{fig:mask_proj}
    \end{subfigure}
    \caption{Suitable segmentation representation to recognize the lateral standard view of the knee: In an ideal standard view the condyles' overlap each other.}
\end{figure}
 \textbf{(1) Field-of-view cropping}: Prevents superposition of the other laterality in the projection domain. \textbf{(2) Laterality alignment}: Prohibits ambiguities during pose estimation. \textbf{(3) Definition of 3D reference planes}: Two independent raters defined the 3D reference planes in the CT volumes utilizing a DRR preview integrated into the open-source Medical Imaging Toolkit \cite{wolf2005medical} with interactive plane positioning. They serve as ground truth pose reference during simulation. \textbf{(4) 3D automatic bone segmentation}: To compute automatic 3D segmentations, a 3D nnU-Net \cite{isensee2021nnu} is trained on a subset of 10 manually annotated CTs for the task of multilabel bone segmentation, segmenting the femur, tibia, fibula, and patella. \textbf{(5) Suitable segmentation representation}: The two femur condyles are not distinguishable in the shape-based  representation, however, this is relevant for optimal lateral view recognition. Annotating the condyles as line features would result in an increased manual labeling effort in the projection domain. Alternatively, we propose to incorporate this information in the segmentation by separating the femur annotation symmetrically along the femoral shaft (Fig.\,\ref{fig:seg_repr}). This results in one additional segmentation label for the lateral standard to recognize condyles' congruence and derive the directional pose offset. \textbf{(6) DRR and mask simulation}: DRRs are simulated for varying angulations of orbital and angular rotation $\alpha,\,\beta\in[-40\degree,40\degree]$ around the defined reference standard. The DeepDRR simulation framework was extended to allow the forward projection of corresponding masks (Fig\,\ref{fig:mask_proj}).
\begin{figure}[t]
    \centering
    \begin{subfigure}{0.17\textwidth}
        \centering
        \includegraphics[width=\textwidth]{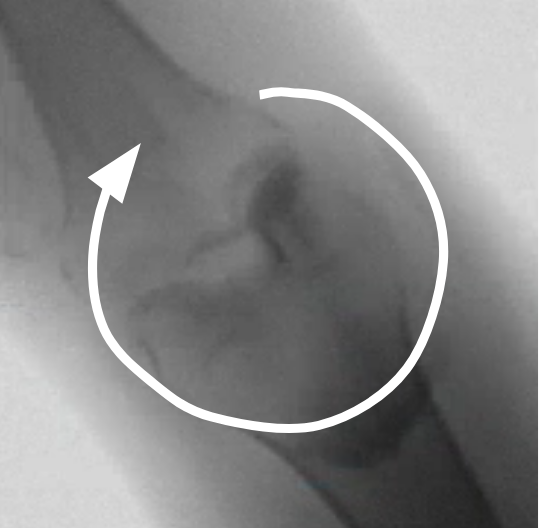}
        \caption{In-plane rotation $\gamma$}
    \end{subfigure}
    \hfill
    \begin{subfigure}{0.17\textwidth}
        \centering
        \includegraphics[width=\textwidth]{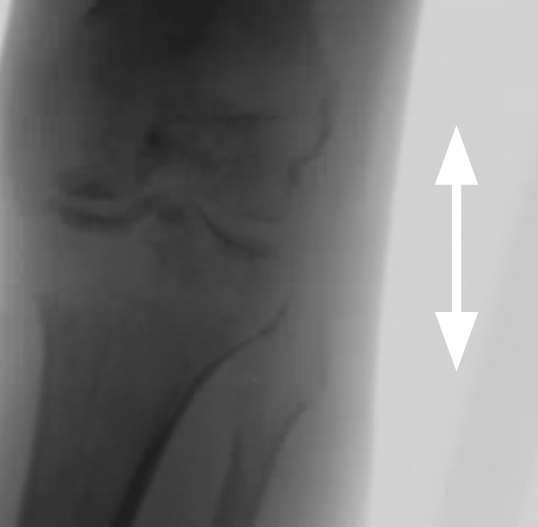}
        \caption{Detector translation}
    \end{subfigure}
    \hfill
    \begin{subfigure}{0.17\textwidth}
        \centering
        \includegraphics[width=\textwidth]{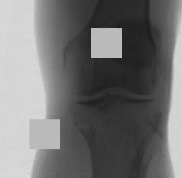}
        \caption{Random region dropout}
    \end{subfigure}
    \hfill
    \begin{subfigure}{0.17\textwidth}
        \centering
        \includegraphics[width=\textwidth]{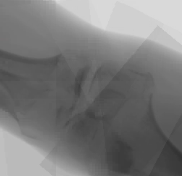}
        \caption{Transparent edges}
    \end{subfigure}
    \hfill
    \begin{subfigure}{0.17\textwidth}
        \centering
        \includegraphics[width=\textwidth]{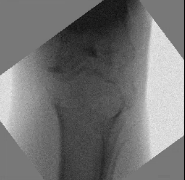}
        \caption{Border overlays}
    \end{subfigure}
    \caption{Training data augmentations.}
    \label{fig:augmentations}
\end{figure}
A set of augmentations is applied to the simulated dataset, to bridge the domain gap from synthetic data to real X-rays (Fig.\,\ref{fig:augmentations}). The in-plane rotation augmentation accounts for variable patient to C-arm alignment, the translation bridges the gap to the validation data where the joint gap is not centered like in the training data, random region dropout reflects superposition artifacts, transparent edge overlays reflect projection artifacts resulting from, e.g., the operating table, and border overlays account for the gamma correction interpolation artifacts.\\
For simulation, a set of 24 CTs was considered, 15 CTs without metal and 9 C-arm volumes with metal. The data was divided $60-20-20\,\%$ for training (15 CTs), validation (5 CTs), test (4 CTs).

\subsection{Shape-based positioning framework:}
The proposed shape-based positioning framework was trained jointly for both standard views (Fig.\,\ref{fig:overview}). It consists of two modules: The first is responsible for a view classification, in-plane rotation and laterality alignment, directly estimated from the image intensities. Thereby, pose ambiguities and data variation are addressed, simplifying the task for the subsequent module to estimate the optimal C-arm positioning for the desired standard view, employing shape features.\\
\textbf{(1) Intensity-based multi-task classification and regression module:} For simultaneous in-plane rotation regression, view recognition, and laterality classification, an EfficientNet-B0 feature extractor \cite{tan2019efficientnet,mairhofer2021ai} was extended with two binary classification heads with one output neuron, followed by sigmoid activation, and one regression head, with the same architecture, but 2 outputs, omitting the activation. The in-plane rotation $\gamma$ is mapped to sin/cos-space to ensure a continuous Loss function during optimization. All training examples were aligned with the same laterality during data simulation which would otherwise result in pose ambiguities. Thus, to train the laterality classifier, the training examples were randomly flipped horizontally with $p=0.5$ and the corresponding $\gamma$ label was adapted accordingly. The weights were optimized using Binary Cross Entropy Loss for the classification heads, and Mean Squared Error for the regression head.\\
\textbf{(2) Shape-based pose regression:} Following surgical characteristics for recognizing correct standard views of the knee, a view-independent shape-based pose regression framework was developed. The architecture is based on a 2D U-Net \cite{klein_mic-dkfzbasic_unet_example_2019} with two view-specific segmentation heads, because the segmentation labels differ for both views. The extracted shape features are used as input for the pose regression network that outputs the necessary C-arm pose update ($\alpha$, $\beta$, $\gamma$, $\mathbf{t}$)$\in\mathbf{R}^6$ to acquire the desired standard view \cite{kausch2020toward}.


\subsection{Validation data:}
Real X-rays for validation were sampled from single Siemens Cios Spin\textsuperscript{\textregistered} sequences generated during 3D acquisition of 6 knee cadavers. Preprocessing consisted of (1) definition of 3D standard reference planes, (2) laterality check, (3) sampling of X-rays around the defined reference standards in the interval $\alpha,\,\beta\in[-30\degree,30\degree]$, and generation of ground truth pose labels. Since the Spin sequences are orbital acquisition sequences, only the orbital rotation is equidistantly covered in the validation set, while the angular rotation is constant for all X-rays sampled from the same sequence. The number of sampled X-rays per standard and view may differ, if the reference standard is located close to the edge of the orbital sequence (range: 102-124).


\section{Experiments and Results}

The proposed pipeline was evaluated considering the following research questions:
\begin{enumerate}
    \item[(RQ1)] Does the proposed shape-based pipeline outperform view-specific intensity-based and shape-based pose regression? How does it influence the generalization from synthetic to real data? (Sec.\,\ref{rq2})
    \item[(RQ2)] How do individual bones influence the overall positioning performance? (Sec.\,\ref{rq3})
    \item[(RQ3)] How accurate is the performance of the view and laterality classification? (Sec.\,\ref{rq4})
\end{enumerate}
Positioning performance was evaluated based on the angle $\theta=\arccos\left(\langle v_{pred},v_{gt} \rangle\right)$ between the principal rays of the ground truth $v_{gt}$ and predicted pose $v_{pred}$ and the mean absolute error ($AE$) of in-plane rotation $\gamma$. The interrater variation of the reference standard planes defined by two independent raters serves as an upper bound for the reachable accuracy of a C-arm positioning approach trained on the reference annotations. It was assessed in terms of orientation differences $\theta$ ($\theta_{a.-p.}=4.1\pm2.6\degree$, $\theta_{lateral}=1.8\pm1.3\degree$).\\
The models were implemented using PyTorch 1.6.0, trained end-to-end with an 11 GB GeForce RTX 2080 Ti, and optimized with the Adam optimizer with a base learning rate of $\eta=10^{-4}$ and batchsize 8 until convergence.


\subsection{Importance of pipeline design choices (RQ1)}
\label{rq2}

\begin{figure}[t]
\centering
\includegraphics[width=0.8\textwidth]{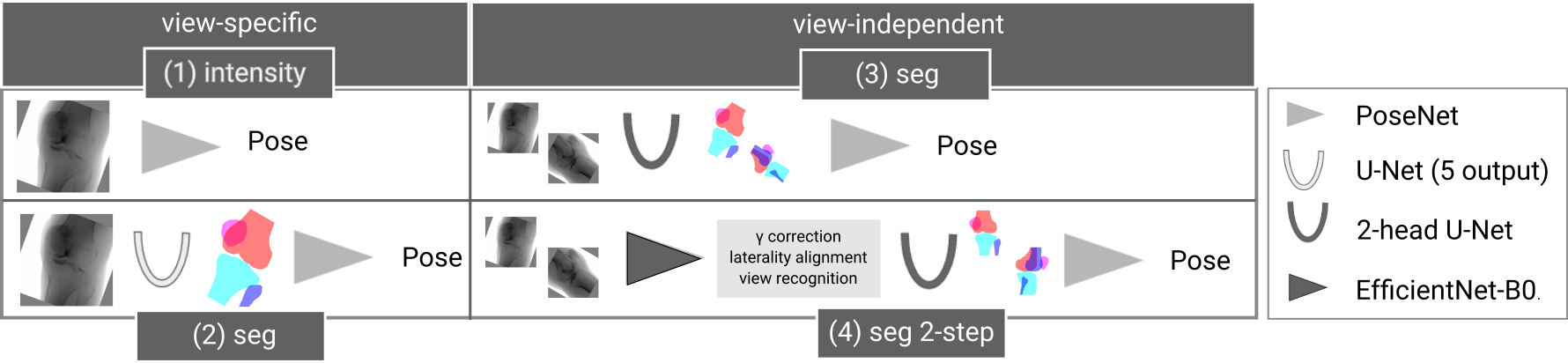}
\caption{Variants for performance comparison.}
\label{fig:variants}
\end{figure}

\begin{figure}[t]
    \centering
    \includegraphics[width=\textwidth]{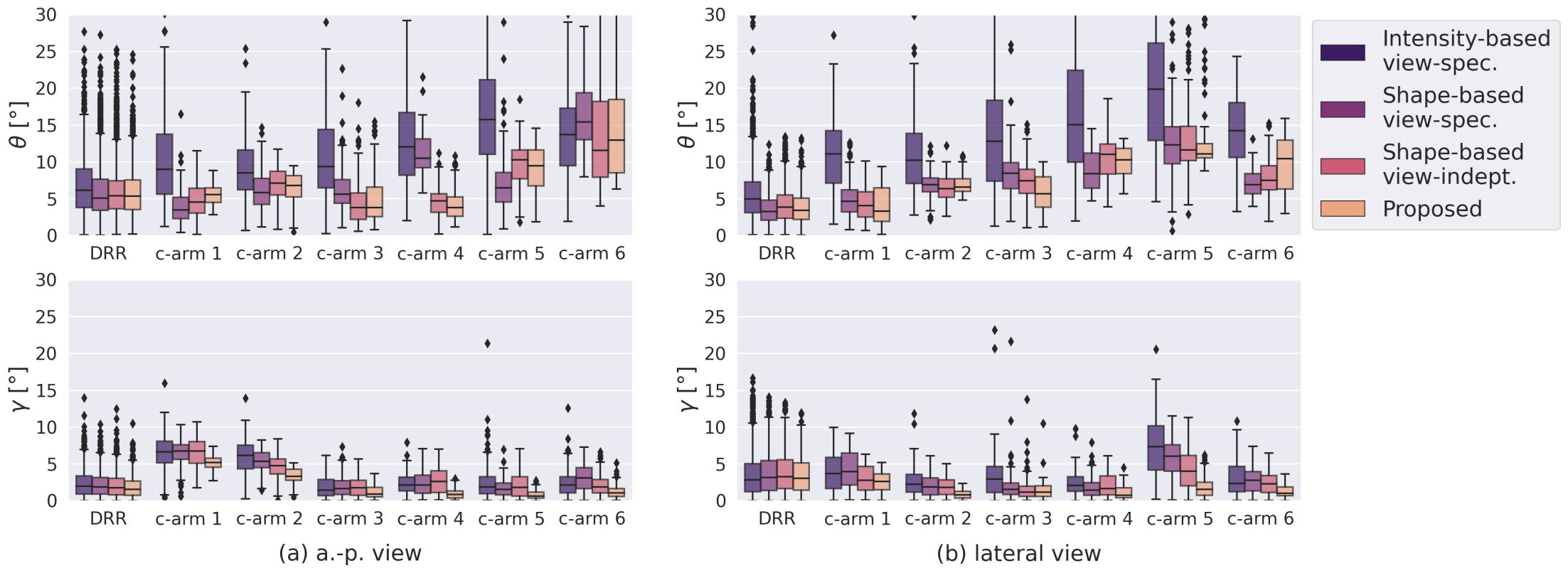}
    \caption{Pose regression performance on simulated and real X-rays compared for different pipeline variants.}
    \label{fig:pose_regression_performance}
\end{figure}
\begin{figure}[t]
    \centering
    \begin{subfigure}{0.503\textwidth}
        \centering
        \includegraphics[width=\textwidth]{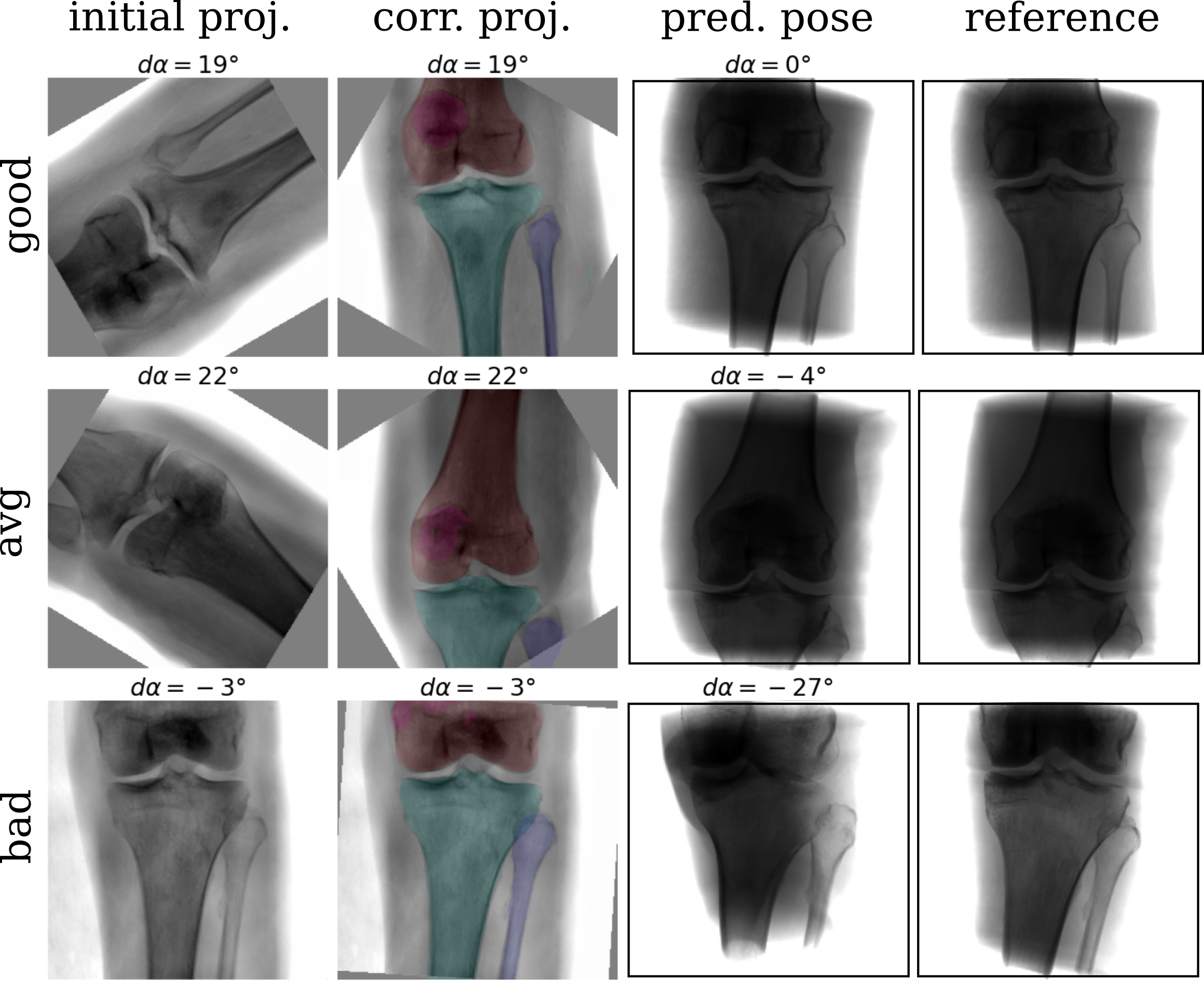}
        \caption{a.-p. view}
    \end{subfigure}
    \hfill
    \begin{subfigure}{0.477\textwidth}
        \centering
        \includegraphics[width=\textwidth]{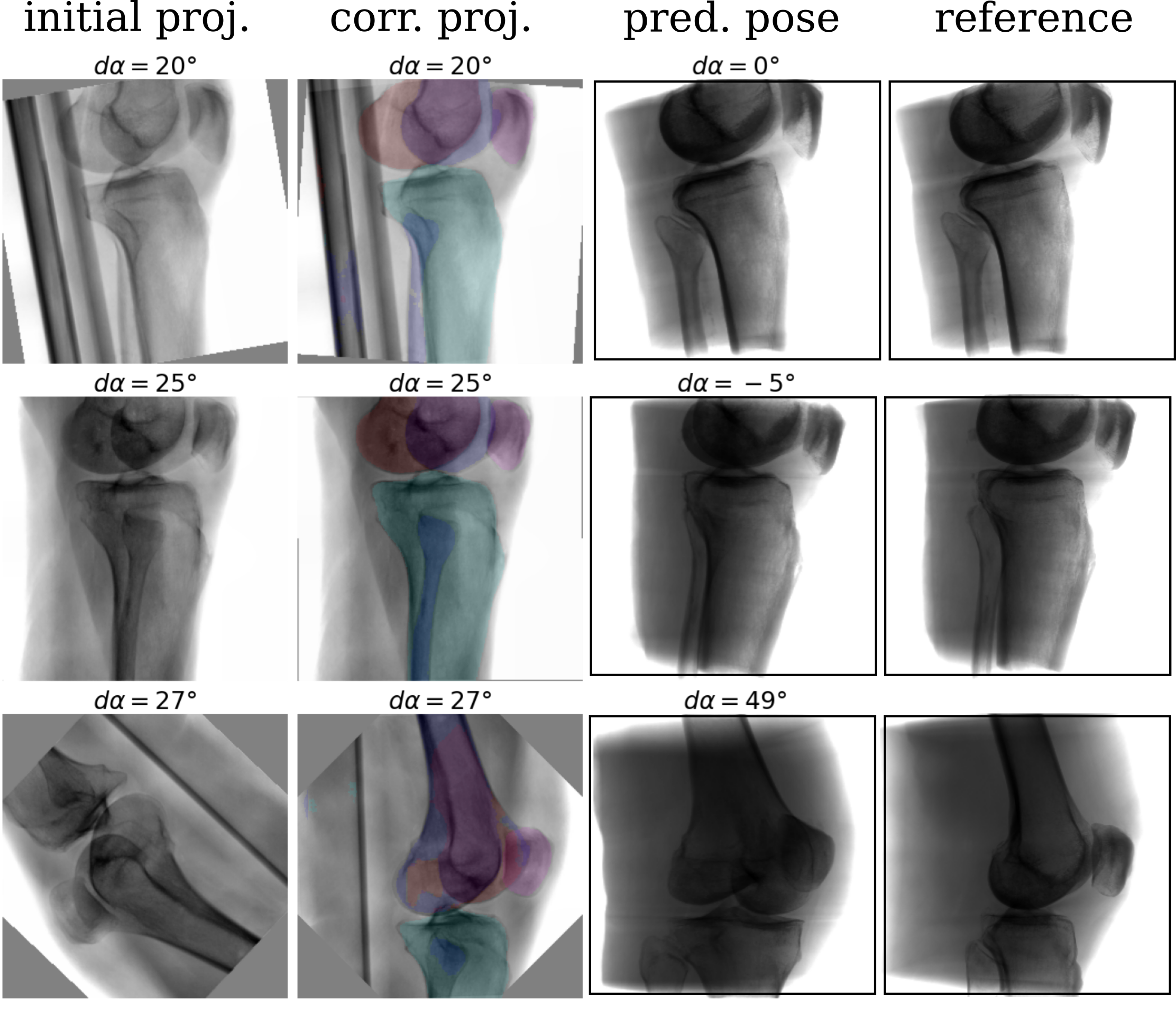}
        \caption{lateral view}
    \end{subfigure}
    \caption{Exemplary visual results for the a.-p. and lateral standard positioning with good, average, and bad performance measured with respect to the orbital rotation offset $\alpha$ to the reference standard pose. The initial real X-ray projection is visualized along with the in-plane corrected projection with the overlaid segmentation mask, and the predicted standard view simulated after C-arm repositioning according to the predicted correction pose side-by-side with the desired reference standard.}
    \label{fig:interrater}
\end{figure}
In an ablation study, the proposed shape-based view-independent pose regression was compared to view-specific direct intensity-based pose regression \cite{kausch2020toward}. Further, the complete pipeline (2-step) is compared to a 1-step segmentation-based approach trained view-specific and view-independent (Fig.\,\ref{fig:variants}). Evaluation was performed on the simulated test DRRs and cadaveric X-rays (Fig.\,\ref{fig:pose_regression_performance}).\\
\textbf{View-independent vs. view-specific networks:} While the view-specific networks perform significantly better (lateral) or comparable (a.-p.) on the simulated data, the view-independent networks perform significantly better (a.-p.) or comparable (lateral) on the real data.\\
\textbf{1-step vs. 2-step:} The proposed 2-step approach performs significantly better or comparable than a 1-step shape-based pose regression approach on most validation cases (8/12) in viewing direction $\theta$. Regarding the $\gamma$ rotation, the 2-step approach improves performance across all validation cases.\\
\textbf{Generalization from DRR to X-ray:} The shape-based pose regression network combined with joint view-independent training clearly boosts the performance compared to direct intensity-based pose regression from $d\theta_{a.-p}^{X-ray}=12.2\pm6.8\degree,\,d\theta_{lateral}^{X-ray}=14.4\pm7.6\degree$ to $d\theta_{a.-p}^{X-ray}=7.4\pm5.0\degree,\,d\theta_{lateral}^{X-ray}=8.4\pm5.4\degree$.


\subsection{Importance of individual bones on overall performance (RQ2)}
\label{rq3}

\begin{figure}[t]
    \centering
    \begin{subfigure}{0.49\textwidth}
        \centering
        \includegraphics[width=\textwidth]{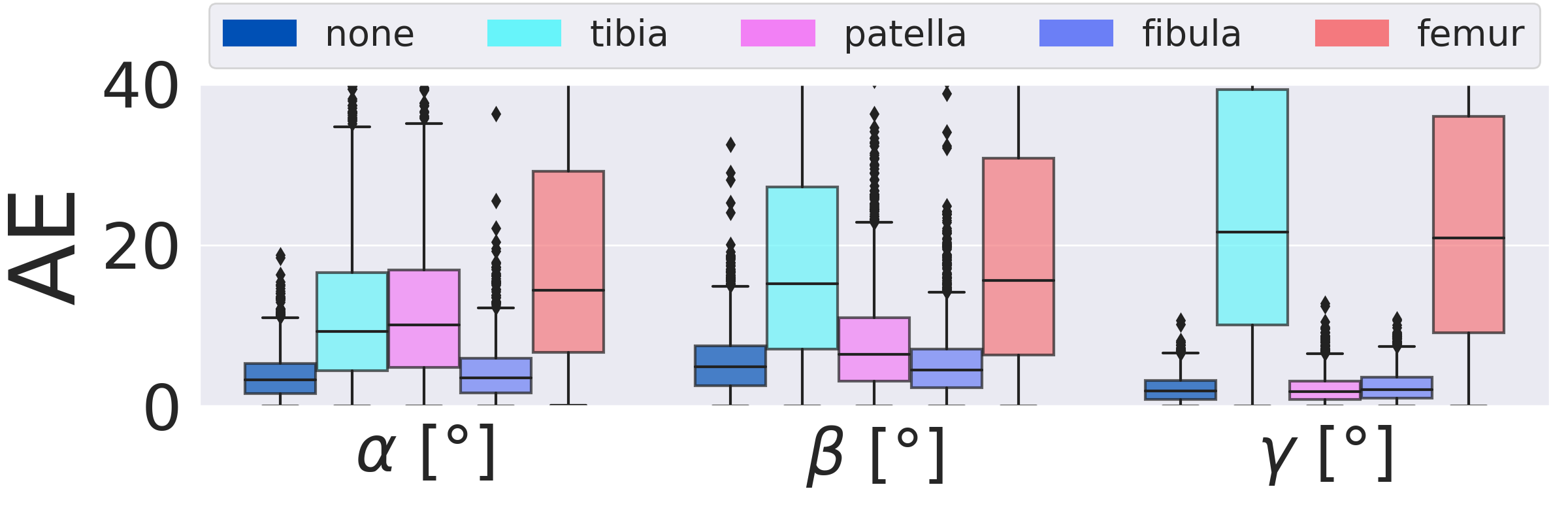}
        \caption{a.-p. view}
    \end{subfigure}
    \hfill
    \begin{subfigure}{0.49\textwidth}
        \centering
        \includegraphics[width=\textwidth]{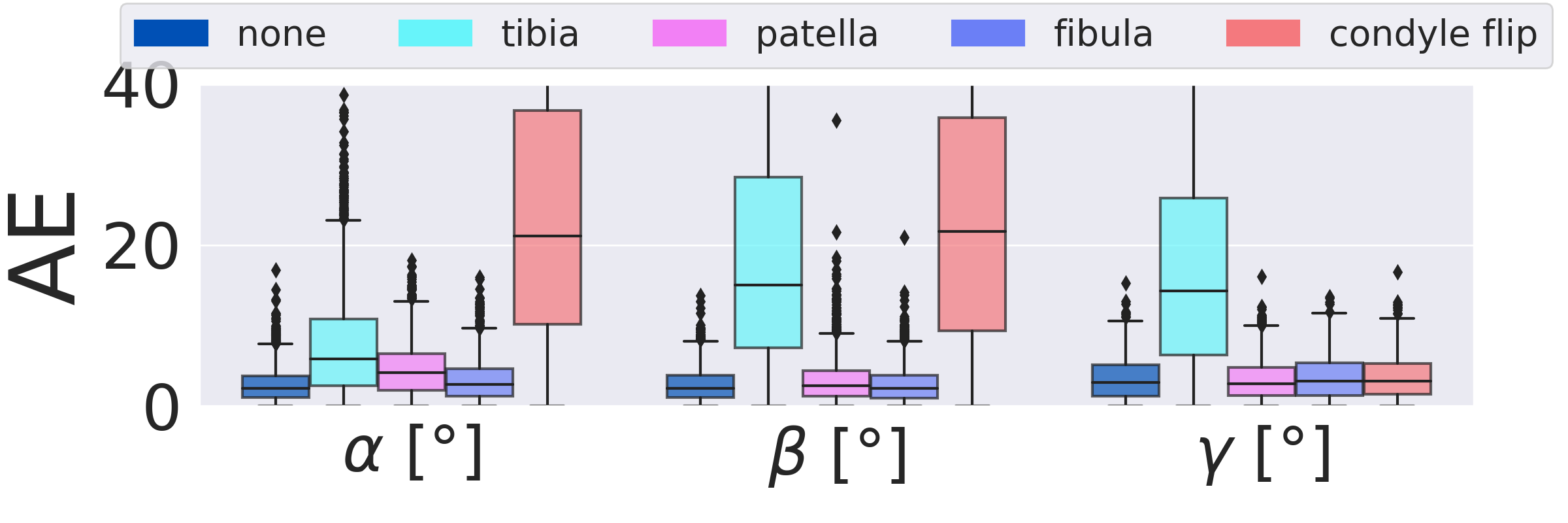}
        \caption{lateral view}
    \end{subfigure}
    \caption{Importance of individual bones on positioning performance (3528 test DRRs). One segmentation channel was set to zero at a time during inference, 'none' corresponds to the reference performance utilizing all channels.}
    \label{fig:dropout_bone}
\end{figure}
Fig.\,\ref{fig:dropout_bone} shows the importance of individual segmented bone classes on the overall positioning performance evaluated on the test DRRs (3528 DRRs). The fibula has very little influence on the positioning for both views. The patella is only important for the a.-p. view, while tibia and femur are relevant for both views. The condyle assignment for the lateral view determines the rotation direction for the orbital and angular rotation ($\alpha$, $\beta$). Inverting the assignment of left and right femur condyle results in a sign flip in $\alpha$, $\beta$.

\subsection{Accuracy of view and laterality classification (RQ3)}
\label{rq4}
The classificator performances were assessed on the synthetic (3528 DRRs, 4 CTs) and real data (1386 X-rays, 6 C-arm scans). The view classificator (a.-p. / lateral) achieved an accuracy of $100\%$ on the test DRRs and $99\%$ on the X-rays. The laterality classificator (left / right) resulted in an accuracy of $98\%$ on the test DRRs and $98\%$ on the X-rays.


\section{Discussion and Conclusion}
A complete framework for automatic acquisition of standard views of the knee is proposed that can handle several standard views simultaneously. The complete pipeline is trained on simulated data with automatically generated annotations and evaluated on real intraoperative X-rays. To bridge the domain gap, different augmentation strategies are suggested that address intraoperative confounding factors, e.g., the OR table. View-independent training and multi-label shape features improve the generalization from simulated training to real X-rays and outperform direct intensity-based approaches. 
View-independent networks result in more training data which showed to improve the generalization from simulated training to real X-rays. The 2-step approach increases robustness and simultaneously automates necessary preprocessing tasks like laterality and standard view recognition, which can be performed with very high accuracy on simulated ($100\%$, $98\%$) and real data ($99\%$, $98\%$). The approach is fast and easy to translate into the operating room as it does not require any additional technical equipment. Assuming that the surgeon acquires the initial X-ray with a pose offset within the capture range of $[-30\degree,\,30\degree]$, it has the potential to reduce time and unnecessary radiation during manual C-arm positioning. Furthermore, the segmentation features can serve as a sanity check and indicate the reliability of the pose regression result. Further experiments with a larger training set covering more anatomical variation, e.g., patella baja and different flexion angles \cite{kronke2022cnn}, can potentially address observed failure cases. 
\tiny
\textbf{Data use declaration:}
The data was obtained retrospectively from anonymized databases and not generated intentionally for the study. The acquisition of data from living patients had a medical indication and informed consent was not required. The corresponding consent for body donation for these purposes has been obtained.
\bibliographystyle{splncs04}
\bibliography{refs}
%
\end{document}